# Scalar Compandor Design Based on Optimal Compressor Function Approximating by Spline Functions

Zoran H. Perić[1], Lazar Z. Velimirović[2*], Miomir S. Stanković[3],

Aleksandra Ž. Jovanović[1], Dragan Antić[1]

[1]Faculty of Electronic Engineering Niš, Aleksandra Medvedeva 14, 18000 Niš, Serbia,

E-mail: zoran.peric@elfak.ni.ac.rs, aleksandra.jovanovic@elfak.ni.ac.rs,

dragan.antic@elfak.ni.ac.rs

[2]Mathematical Institute of the Serbian Academy of Sciences and Arts, Kneza Mihaila

36, 11001 Belgrade, Serbia, E-mail: lazar.velimirovic@mi.sanu.ac.rs,

Phone: +381 (18) 529-101, Fax: +381 (18) 588-399

[3]Faculty of Occupational Safety Niš, Čarnojevića 10 A, 18000 Niš, Serbia,

E-mail: miomir.stankovic@znrfak.ni.ac.rs

Abstract — In this paper the approximation of the optimal compressor function using the first-degree spline functions and quadratic spline functions is done. Coefficients on which we form approximative spline functions are determined by solving equation systems that are formed from treshold conditions. For Gaussian source at the input of the quantizer, using the obtained approximate spline functions a companding quantizer designing is done. On the basis of the comparison with the SQNR of the optimal compandor it can be noticed that the proposed companding quantizer based on approximate spline functions achieved SQNR arbitrary close to that of the optimal compandor.

---

[*]Corresponding author

## I. Introduction

Quantization is the process of approximating continuous-amplitude signals by discrete-amplitude signals and it is an important aspect of data compression or coding. The independent quantization of each signal value or parameter is termed scalar quantization. A device that performs quantization is called a quantizer [1]. When all the quantization intervals are of the same width, the quantizer is uniform. Otherwise, the quantizer is nonuniform (different width of the quantization intervals). Uniform quantizers are suitable for signals that have approximately uniform distribution [1]. How most of the signals do not have a uniform distribution there is a need for using nonuniform quantizer.

Nonuniform quantization can be realized using the companding technique. Namely nonuniform quantization can be achieved by compressing the signal $x$ using a nonuniform compressor characteristic $c(\cdot)$, by quantizing the compressed signal $c(x)$ employing a uniform quantizer, and by expanding the quantized version of the compressed signal using a nonuniform transfer characteristic $c^{-1}(\cdot)$ that is inverse to that of the compressor. The overall structure of a nonuniform quantizer consisting of a compressor, a uniform quantizer, and expandor in cascade is called compandor [1].

Quantizer is completely defined if nonlinear compressor function is known. However, it is well known that designing the optimal nonlinear companding quantizers for a Gaussian source is very complex due to the difficulties in determining the inverse optimal compressor function [1-5]. In order to simplify realization of the optimal nonlinear compressor function, its approximation by other function is often done. This way the quantizer design is considerably simplified.

By applying the similar methodology as in [6], in this paper we propose design of companding quantizer of a Gaussian source by approximating the optimal compressor function using the first-degree spline functions and quadratic spline functions. The support region of companding quantizer is divided into $2L = 4$ equal segments. The number of cells within segments is different. With such quantizer design, the value of the signal to quantization noise ratio SQNR that is very close to the value of SQNR of the optimal compandor is achieved.

The remainder of the paper is organized as follows: In section II the detailed description of the first-degree spline functions and quadratic spline functions, as well as their application in approximation of the optimal compressor function for Gaussian source are given. Scalar compandor design based on optimal compressor function approximating by spline functions is described in section III. Finally, section IV presents a discussion of numerical results obtained for the SQNR of the proposed companding quantizer for the Gaussian source of unit variance.

## II. Approximations of the Optimal Compressor Function Using Spline Functions

In this paper, the approximation of the optimal compressor function using the first-degree spline functions and the quadratic spline functions is done. A spline function is a function that consists of polynomial pieces joined together with certain smoothness conditions. The general definition of spline functions of arbitrary degree is as follows [7].

Definition. A function $S$ is called a spline of degree $k$ if:

1. The domain of $S$ is an interval $[a, b]$.

2. $S, S',..., S^{(k-1)}$ are all continuous functions on interval $[a, b]$.

3. There are points $x_i$ (the knots of $S$) such that $a = x_0 < x_1 < ... < x_L = b$ and such that $S$ is a polynomial of degree at most $k$ on each subinterval $[x_i, x_{i+1}]$.

The optimal compressor function $c(x)$: $[-x_{\max}, x_{\max}] \to [-x_{\max}, x_{\max}]$, by which the maximum of SQNR is achieved for the reference variance $\sigma^2$ of an input signal $x$ is defined as [7]:

$$c(x) = \begin{cases} x_{\max} \dfrac{\int_0^{|x|} p^{1/3}(t)dt}{\int_0^{x_{\max}} p^{1/3}(t)dt} \operatorname{sgn}(x), & |x| \le x_{\max}, \end{cases} \quad (1)$$

where $x_{\max}$ denotes the support region threshold of the optimal companding quantizer and $p(t)$ is a symmetric Gaussian probability density function (PDF), which is defined as follows [1]:

$$p(t) = \frac{1}{\sqrt{2\pi}\sigma} \exp\left(-\frac{t^2}{2\sigma^2}\right). \quad (2)$$

By substituting (2) into (1) the well known form of the optimal compressor function of a Gaussian source is obtained:

$$c(x) = \begin{cases} x_{\max} \dfrac{1 - \operatorname{erfc}\left(\dfrac{|x|}{\sqrt{6}\sigma}\right)}{1 - \operatorname{erfc}\left(\dfrac{x_{\max}}{\sqrt{6}\sigma}\right)} \operatorname{sgn}(x), & |x| \le x_{\max}, \end{cases} \quad (3)$$

where $\operatorname{erfc}(u) = \dfrac{2}{\sqrt{\pi}} \int_u^\infty \exp(-t^2)dt$. Since $c(-x) = -c(x)$ holds, in continuation of the paper the approximation of the optimal compressor function is constrained on domain $[0, x_{\max}]$.

The polygonal function is the first-degree spline function which consists of linear polynomials joined together to achieve continuity [7]. According to [7] we have:

$$S(x) = \begin{cases} S_0(x), & x \in [x_0, x_1] \\ S_1(x), & x \in [x_1, x_2] \\ \vdots \\ S_{L-1}(x), & x \in [x_{L-1}, x_L] \end{cases}, \qquad (4)$$

where

$$S_i(x) = a_i x + b_i. \qquad (5)$$

Obviously, $S(x)$ is a piecewise linear function. If the function $S(x)$ defined by (4) is continuous, we call it a first-degree spline.

The first-degree spline function, $g^{s1}(x)$, by which the optimal nonlinear compressor function $c(x)$ is approximated in this paper, for the number of segments in the positive part of the axis $L = 2$, has the following form [7]:

$$g^{s1}(x) = \begin{cases} c_1(x_1) + m_1(x - x_1), & x \in [0, x_1] \\ c_2(x_2) + m_2(x - x_2), & x \in [x_1, x_2] \end{cases}, \qquad (6)$$

The coefficients of direction of the line, $m_1$ and $m_2$, are given by the formula:

$$m_i = \frac{c_i(x_i) - c_{i-1}(x_{i-1})}{x_i - x_{i-1}}, \quad i = 1, \ldots, L. \qquad (7)$$

The quadratic spline is consisted of parabola parts between two consecutive nodes, chosen to have the same tangent at node. The approximate quadratic spline function, $g^{s2}(x)$, which approximates a nonlinear compressor function $c(x)$, for the $L = 2$ segments in the positive part of the axis, has the following form [7]:

$$g^{s2}(x) = \begin{cases} a_1 + b_1 x + d_1 x^2, & x \in [0, x_1] \\ a_2 + b_2 x + d_2 x^2, & x \in [x_1, x_2] \end{cases}, \qquad (8)$$

To determine the coefficients of the approximate function $g(x)$, it is necessary to fulfill the defined requirements for the quadratic spline function. The number of set conditions is equal to the number of coefficients that should be determined. In fact, as we have three nodes and two subintervals, and each the second-degree polynomial has three coefficients, means we have to determine the six coefficients in total [7]. One should solve the following system of equations, where the number of set conditions is equal to the number of the coefficients to be determined [7]:

$$g^{s2}(0) = 0, \tag{9}$$

$$c_1(x_1) = a_1 + b_1 x_1 + d_1 x_1^2, \tag{10}$$

$$c_1(x_1) = a_2 + b_2 x_1 + d_2 x_1^2, \tag{11}$$

$$c_2(x_2) = a_2 + b_2 x_2 + d_2 x_2^2, \tag{12}$$

$$\lim_{x \to x_1^-} g^{s2'}(x) = \lim_{x \to x_1^+} g^{s2'}(x) \Rightarrow b_1 + 2d_1 x_1 = b_2 + 2d_2 x_1, \tag{13}$$

$$g^{s2'}(x_2) = 0 \Rightarrow b_2 + 2d_2 x_2 = 0. \tag{14}$$

In such a manner, the coefficient values are obtained for: $a_1, b_1, d_1, a_2, b_2, d_2$.

### III. Design of Scalar Compandor Based on Approximate Spline Functions

This section provides a detailed description of the scalar compandor whose compressor function is the first-degree or quadratic spline approximation of the optimal compressor function for a Gaussian source. Since the following derivation holds for both scalar compandors, due to simplicity, the model notations are omitted. The support region threshold for the optimal companding quantizer is defined as follows [3]:

$$x_{max} = \sigma \sqrt{6 \ln N} \left[ 1 - \frac{\ln \ln N}{4 \ln N} - \frac{\ln(3\sqrt{\pi})}{2 \ln N} \right]. \tag{15}$$

In our quantizer design, the total number of the reproduction levels per segments in the first quadrant is assumed to be $(N\text{-}2)/2$:

$$\sum_{i=1}^{L} \frac{N_i}{2} = \frac{N-2}{2}, \quad (16)$$

where the number of reproduction levels per segments, $N_i/2$, is determined from the condition that all reproduction levels are equidistant at compressor output:

$$\frac{N_i}{2} = \frac{N-2}{2} \frac{c_i(x_i) - c_{i-1}(x_{i-1})}{c_L(x_L)}, \quad i = 1, \ldots, L. \quad (17)$$

Reproduction levels of the companding quantizer we propose are determined using the inverse spline approximation of $c(x)$ as follows:

$$y_{i,j} = g_i^{-1}\left(\left(\frac{2j-1}{2}\right)\Delta\right), \quad i = 1, \; j = 1, \ldots, \frac{N_i}{2}, \quad (18)$$

$$y_{i,j} = g_i^{-1}\left(c_i(x_i) + \left(\frac{2j-1}{2}\right)\Delta\right), \quad i = 2, \ldots, L, \; j = 1, \ldots, \frac{N_i}{2}, \quad (19)$$

where for the $y_{i,j}$ is taken the solution that belongs to the spline function domain. Indexes $i$ and $j$ indicate the $j$-th reproduction levels within the $i$-th segment. Step size at the output of compressor $\Delta$ is determined with:

$$\Delta = \frac{2x_{\max}}{N-2}, \quad (20)$$

while the $j$-th cell length within the $i$-th segment of the considered companding quantizer is determined by:

$$\Delta_{i,j} = \frac{\Delta}{g_i'(y_{i,j})}, \quad i = 1, \ldots, L, \; j = 1, \ldots, \frac{N_i}{2}. \quad (21)$$

The total distortion is a quality measure of quantization process, and can be found as a sum of the granular $D_g$ and the overload $D_o$ distortion. The granular distortion for a compandor is defined with Benett's integral [1], which in the case of segmental compressor function gets form:

$$D_g = 2 \frac{x_{\max}^2}{3(N-2)^2} \sum_{i=1}^{L} \sum_{j=1}^{N_i/2} \frac{p(y_{i,j})}{[g_i'(y_{i,j})]^2} \Delta_{i,j}. \quad (22)$$

The overload distortion is given by [1]:

$$D_o = 2 \int_{x_{max}}^{\infty} (x - y_{max})^2 p(x) dx, \qquad (23)$$

where the $y_{max}$ is the last reproduction level determined from the centroid condition [1]:

$$y_{max} = \frac{\int_{x_{max}}^{\infty} x p(x) dx}{\int_{x_{max}}^{\infty} p(x) dx}. \qquad (24)$$

Finally, combining the set of equations (15), (23) and (24), for the Gaussian PDF (2), one can derive the overload distortion of the proposed companding quantizer as in [3]:

$$D_o = \sqrt{\frac{2}{\pi}} \frac{1}{x_{max}^3} \exp\left(-\frac{x_{max}^2}{2}\right). \qquad (25)$$

By determining the total distortion $D$, the signal to quantization noise ratio can be determined [1]:

$$\text{SQNR} = 10 \log\left(\frac{\sigma^2}{D_g + D_o}\right) = 10 \log\left(\frac{\sigma^2}{D}\right). \qquad (26)$$

Without losing the generality, in what follows the quantizer design will be done for the reference input variance of $\sigma^2 = 1$.

### IV. Numerical Results and Conclusions

Numerical results presented in this section are obtained for the case when the number of segments is equal to $2L = 4$ and for different values of the number of levels $N$. Table 1 shows values of the segment threshold $x_1$, values of the support region threshold $x_2 = x_{max}$, values of the optimal compressor functions $c(x_1)$ and $c(x_2)$ and values of the coefficient of direction of the line, $m_1$ and $m_2$, for a different number of levels $N$ (from $N = 16$ to $N = 128$). Based on these parameters the first-degree spline approximation is formed. Substituting the appropriate values of the parameters from Table 1 in (6), we obtain the first-degree spline approximation.

Table 2 shows the values of the coefficients which are determined by (9) - (14) based on which an approximate quadratic spline function is formed. Substituting the appropriate values of the coefficients from Table 2 in (8), we obtain the approximate quadratic spline function.

In Fig. 1 the dependence of the optimal compression function, $c(x)$, the first-degree spline approximation, $g^{s1}(x)$, and the quadratic spline approximation, $g^{s2}(x)$, on input signal $x$ for the number of levels $N = 128$ is shown. Based on Fig. 1 it can be concluded that the quadratic spline better approximates the optimal compressor function than the first-degree spline.

Table 3 shows the values of SQNR of the quantizer described in [2], (SQNR$^{RS}$), the values of SQNR of the proposed companding quantizer based on the first-degree spline approximation, (SQNR$^{FDS}$), the values of SQNR of the proposed companding quantizer based on the quadratic spline approximation, (SQNR$^{QS}$) and the values of SQNR of quantizer based on the optimal compression function $c(x)$, (SQNR$^{OC}$), for a different number of levels $N$. In [2], a piecewise uniform scalar quantizer of a Gaussian source is designed assuming the equidistant thresholds of segments, while the number of reproduction levels inside segments is determined by optimizing the granular distortion.

Fig. 2 presents the dependency of SQNR on the number of bits per sample for the quantizer described in [2], the proposed companding quantizer based on approximate first-degree spline functions, the proposed companding quantizer based on quadratic spline functions and the optimal compandor.

Analyzing the results shown in Fig. 2 and Table 3 it can be noticed that the SQNR value of the proposed companding quantizer based on the first-degree spline approximation is very close to the SQNR value of the quantizer described in [2]. As the proposed companding quantizer based on the quadratic spline approximation achieves much higher SQNR than the proposed companding quantizer based on the first-degree spline approximation, it can be concluded that the application of the quadratic spline approximation, for the same number of segments and the number of levels, provides much higher SQNR than the quantizer described in [2]. One of conclusion, that can be also observed, is that with the increase of $N$ the $SQNR^{QS}$ approaches to the $SQNR^{OC}$, which isn't the case with the $SQNR^{FDS}$. Thus, the scalar compandor of a Gaussian source whose compressor function is the quadratic spline approximation of the optimal compressor function can achieve SQNR arbitrary close to that of the optimal compandor. Although the compressor function of this quantizer consists of parabola parts, it is easily to find the inverse compressor function, which is the main difficulty with the optimal compression function of a Gaussian source [1]. Withal the number of segments is small (4). It can be summarized that the proposed model is a very effective solution because it represents a simple quantizer model that achieves a high quality signal.

## Acknowledgment

This work is partially supported by Serbian Ministry of Education and Science through Mathematical Institute of Serbian Academy of Sciences and Arts (Project III44006) and by Serbian Ministry of Education and Science (Project TR32035).

**Table caption:**

Table 1. The values of the parameters on which the first-degree spline function is formed

| N | $x_1$ | $x_2 = x_{max}$ | $c(x_1)$ | $c(x_2)$ | $m_1$ | $m_2$ |
|---|-------|-----------------|----------|----------|-------|-------|
| 16 | 1.2373 | 2.4746 | 1.5339 | 2.4746 | 1.2397 | 0.7603 |
| 32 | 1.5259 | 3.0519 | 2.0579 | 3.0519 | 1.3487 | 0.6514 |
| 64 | 1.7819 | 3.5638 | 2.5843 | 3.5638 | 1.4503 | 0.5497 |
| 128 | 2.0137 | 4.0274 | 3.1029 | 4.0274 | 1.5409 | 0.4591 |

Table caption:

**Table 2.** The values of the coefficients on which the quadratic spline function is formed

| N | $x_1$ | $x_2 = x_{max}$ | $a_1$ | $b_1$ | $d_1$ | $a_2$ | $b_2$ | $d_2$ |
|---|-------|-----------------|-------|-------|-------|-------|-------|-------|
| 16 | 1.2373 | 2.4746 | 0 | 0.9588 | 0.2269 | 1.2882 | 3.0411 | 0.6144 |
| 32 | 1.5259 | 3.0519 | 0 | 1.3945 | 0.0301 | 0.9238 | 2.6054 | 0.4269 |
| 64 | 1.7819 | 3.5638 | 0 | 1.8012 | 0.1969 | 0.3542 | 2.1988 | 0.3085 |
| 128 | 2.0137 | 4.0274 | 0 | 2.1636 | 0.3092 | 0.3294 | 1.8364 | 0.2279 |

**Table caption:**

**Table 3.** The values of SQNR for the proposed companding quantizers and the optimal compandor

| N | SQNR$^{RS}$ [dB] | SQNR$^{FDS}$ [dB] | SQNR$^{QS}$ [dB] | SQNR$^{OC}$ [dB] |
|---|---|---|---|---|
| 16 | 19.36 | 19.51 | 19.69 | 20.22 |
| 32 | 25.33 | 25.35 | 25.80 | 26.01 |
| 64 | 31.08 | 31.07 | 31.88 | 31.89 |
| 128 | 36.82 | 36.74 | 37.80 | 37.81 |

**Figure caption:**

**Fig. 1** The first-degree spline function, the quadratic spline function and nonlinear optimal compressor function for the number of segments $2L = 4$ and the number of levels $N = 128$

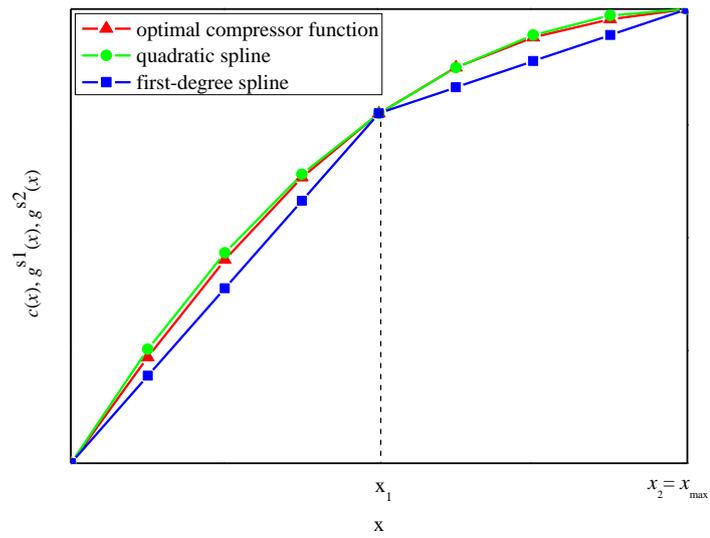

**Figure caption:**

**Fig. 2** Dependency of SQNR on the number of bits per sample for the quantizer described by [2], the proposed companding quantizers and the optimal compandor.for the number of segments $2L = 4$

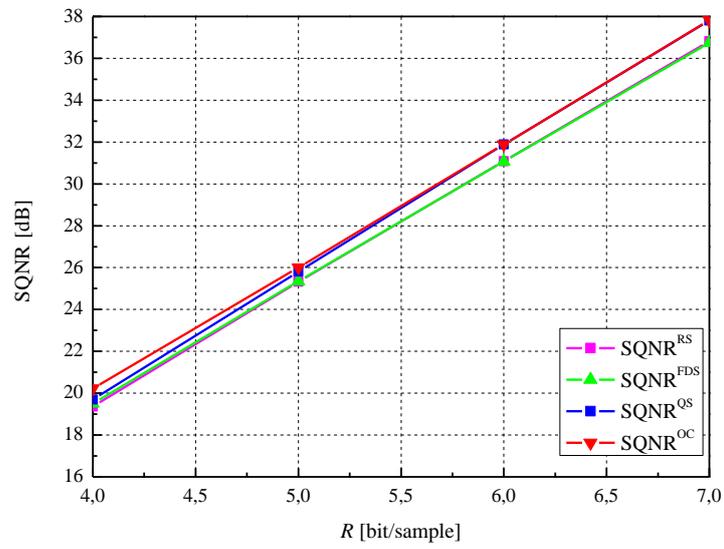